\pdfoutput=1 
\documentclass[11pt]{article}
\usepackage[utf8]{inputenc}
\usepackage[DIV12]{typearea}
\usepackage{amsmath,amsfonts,amssymb,amscd}
\usepackage[all]{xy}
\usepackage{graphicx}
\usepackage{cite}
\usepackage{mathrsfs}
\usepackage{bbm}
\usepackage{bbold}
\usepackage{xcolor}
\usepackage{booktabs}
\usepackage{multirow}
\usepackage[normalem]{ulem}
\usepackage{tensor}
\usepackage{mathtools}
\usepackage[bookmarks=false]{hyperref}
\hypersetup{
    pdftitle = {Goofy Symmetries},
    pdfauthor = {Andreas Trautner},
    colorlinks=true,
    linkcolor=black,
    filecolor=black,      
    urlcolor=black,
    citecolor=black
}
\usepackage{srcltx}
\usepackage{tikz}
\usepackage{mciteplus}

\newcommand{\SU}[1]{\ensuremath{\mathrm{SU}(#1)}}

\newcommand{\U}[1]{\ensuremath{\mathrm{U}(#1)}}
\newcommand{\Z}[1]{\ensuremath{\mathbbm{Z}_{#1}}} 

\newcommand{\rep}[2][]{\ensuremath{\boldsymbol{#2}#1}}

\renewcommand{\bar}[1]{\overline{#1}}

\newcommand{\mat}[1]{\left(\begin{matrix}#1\end{matrix}\right)}

\newcommand{\I}{\ensuremath{\mathrm{i}}}

\definecolor{darkgreen}{HTML}{109930}

\allowdisplaybreaks[1]

\def\mytitle{Goofy is the new Normal}

\title{\mytitle}

\begin{document}

\begin{titlepage}
\setlength{\topmargin}{0.0 true in}
\thispagestyle{empty}

\vspace*{1.0cm}


\begin{center}
{\LARGE\textbf{\mytitle}}

\renewcommand*{\thefootnote}{\fnsymbol{footnote}}

\vspace{1.6cm}
\large
Andreas Trautner
\footnote[1]{
\href{mailto:trautner@cftp.ist.utl.pt}{trautner@cftp.ist.utl.pt}
}
\normalsize
\\[5mm]
\textit{
Max-Planck-Institut f\"ur Kernphysik \\ Saupfercheckweg 1, 69117 Heidelberg, Germany \\[0.2cm]
and \\[0.2cm]
CFTP, Departamento de F\'isica, Instituto Superior T\'ecnico, Universidade de Lisboa \\ Avenida Rovisco Pais 1, 1049 Lisboa, Portugal
}

\end{center}
\vspace*{12mm}

\begin{abstract}\noindent
We discuss the recently discovered global \textit{goofy} transformations and conclude that their understanding 
is mandatory to describe the renormalization group (RG) fixed point structure of quantum field theories (QFTs) in general. Originally, goofy transformations were identified in the two-Higgs-doublet model~(2HDM), where their surprisingly RG stable parameter relations do not correspond to any \textit{regular} symmetry. We unveil several new goofy transformations and their associated RG fixed points, which we have explicitly checked to be RG invariant to three-loop order in gauge-, and scalar quantum corrections. We give a general argument as to why the implied parameter relations are RG invariant to all orders, even though the corresponding goofy transformations are explicitly broken by the gauge-kinetic terms. Goofy transformations can prohibit bare scalar mass parameters and give rise to generation-dependent sensitivity on relative signs of gauge couplings, for what reason they may have important applications for the electroweak hierarchy problem and the Standard Model flavor puzzle. We also discuss the relevance of goofy-symmetric RG fixed points for a phenomenon of ``dynamical classicalization'' in QFT.
\end{abstract}
\thispagestyle{empty}
\clearpage

\end{titlepage}

\widowpenalty10000\clubpenalty10000
\section{Introduction}
Recently, Ferreira, Grzadkowski, Ogreid, and Osland~(FGOO) suggested the presence of new, so far unknown, possible global symmetries of the 2HDM scalar potential~\cite{Ferreira:2023dke}.\footnote{%
We refer to these new supposed symmetries and their associated transformations as \textit{goofy}, as originally suggested by P.~M.~Ferreira~\cite{Ferreira:2023} (in~\cite{Ferreira:2023dke, Grzadkowski:2024} these transformations are referred to as ``$r_0$ transformations'').
We deem this a suitable acronym not only for the surnames of the original authors (FGOO) and the unusual nature of the associated transformations, but also because ``goofy'' has subsequently been elected as the German~\href{https://en.wikipedia.org/wiki/Youth_word_of_the_year_(Germany)}{``Youth word of the year''} 2023~\cite{enwiki:1279583156}.}
This came as a surprise, insofar as the community believed that all possible global symmetries of the 2HDM scalar potential had been discovered and classified~\cite{Ivanov:2005hg,Ivanov:2006yq,Ivanov:2007de,Ferreira:2010yh} (see also~\cite{Deshpande:1977rw,Ginzburg:2004vp,Nishi:2011gc,Battye:2011jj,Pilaftsis:2011ed,Dev:2014yca,Darvishi:2019dbh}).
The suggested new \textit{goofy} symmetries were identified based on the renormalization group (RG) invariance of their
associated coupling relations~\cite{Ferreira:2023dke}. This was explicitly checked using the beta functions at one-loop~\cite{Haber:1993an,Branco:2011iw}
and up to three-loop order~\cite{Bednyakov:2018cmx}, with general arguments based on basis covariants~\cite{Ivanov:2005hg} and their running~\cite{Bednyakov:2018cmx,Bednyakov:2025sri}, suggesting that the relations are indeed stable to all orders~\cite{Ferreira:2023dke}. The new \textit{goofy} symmetries have also been investigated in the full effective potential~\cite{Cao:2023kgq,Pilaftsis:2024uub},
where they could be preserved, if one would allow for an according sign flip of the cutoff~\cite{Grzadkowski:2024}. 
Recently, \textit{goofy} symmetries have also been identified in a two-real-scalar-field toy model~\cite{Grzadkowski:2024}, where they seem to coincide with actual \textit{regular} symmetries of the complexified version of the theory~\cite{Haber:2025cbb}.

In this work, we investigate the nature of goofy transformations and bring their treatment in line with the well-known flavor- and CP-type global symmetries of the 2HDM scalar potential. A non-trivial transformation of the kinetic terms is established as a defining criterion for the goofy nature of a transformation. We uncover several new goofy transformations, which correspond to RG stable parameter hyperplanes of the 2HDM that were not previously known. 

Strictly enforcing a goofy transformation on the kinetic terms would lead to a vanishing of one or several kinetic terms. Hence, none of the goofy symmetries should strictly be conserved as global symmetry in a fully dynamical theory.
Nonetheless, the parameter relations implied by goofy transformations are RG stable to all orders in scalar quantum corrections, and we present a very general argument as to why this is the case. In the simplest cases, even gauge quantum corrections cancel to all orders. In general, non-vanishing RG corrections to goofy parameter relations can arise at higher orders in gauge loops, and we expect them to become sensitive to the \textit{relative} sign of gauge couplings across different generations of flavor. Quantum corrections involving Higgs Yukawa couplings generically break the goofy relations, unless the goofy transformation is enforced on them from the start. 

In the next section we define our notation and review the well-known regular flavor- and CP-type transformations of the 2HDM. In Sec.~\ref{sec:goofy} we recast the original goofy transformation of~\cite{Ferreira:2023dke}. Subsequently, we discuss general field redefinitions and how they transform the kinetic terms. We discuss the physicality of the sign of the kinetic term in the goofy-symmetric theory and show that it can be flipped to the canonical case by a ``goofy basis change'' which commutes with the goofy transformation. 
In Sec.~\ref{sec:generalgoofy} we uncover several new goofy transformations of the 2HDM and their associated parameter relations, summarized in Tab.~\ref{tab:Goofy_relations}.
In Sec.~\ref{sec:goofybeyondsigns} we generalize the goofy transformations beyond global sign flips. Our general argument for the RG stability of goofy parameter relations is given in Sec.~\ref{sec:RGstability}. In Sec.~\ref{sec:noncanonical}, we extend our discussion to non-canonical kinetic terms. We give several general arguments for the importance of goofy symmetries and their continued investigation in Sec.~\ref{sec:comments}.
\section{Flavor- and CP-type transformations}
In a 2HDM, each of the scalar doublets $\Phi_{a=1,2}(x)$ transforms as $(\rep{2},-1/2)$ under the SM gauge group $\SU{2}_\mathrm{L}\times\U1_\mathrm{Y}$. In terms of real degrees of freedom $\phi_{i=1,\dots,8}$ these can be decomposed~as
\begin{equation}\label{eq:realDOF}
\Phi_1:=\mat{\phi_1+\I\phi_2 \\ \phi_3+\I\phi_4}\,,\quad\text{and}\quad
\Phi_2:=\mat{\phi_5+\I\phi_6 \\ \phi_7+\I\phi_8}\,.
\end{equation}
The most general gauge invariant scalar potential can be written as
\begin{equation}
\begin{split}\label{eq:pot}
 V ~&=~ m_{11}^2\Phi_1^\dagger\Phi_1+m_{22}^2\Phi_2^\dagger\Phi_2+\lambda_1(\Phi_1^\dagger\Phi_1)^2
+\lambda_2(\Phi_2^\dagger\Phi_2)^2
+\lambda_3(\Phi_1^\dagger\Phi_1)(\Phi_2^\dagger\Phi_2)
+\lambda_4(\Phi_1^\dagger\Phi_2)(\Phi_2^\dagger\Phi_1) \\
&\qquad-\left\{m_{12}^2\Phi_1^\dagger\Phi_2\right\}
+\left\{\lambda_5(\Phi_1^\dagger\Phi_2)^2
+\big[\lambda_6(\Phi_1^\dagger\Phi_1)
+\lambda_7(\Phi_2^\dagger\Phi_2)\big]
\Phi_1^\dagger\Phi_2\right\}+{\rm h.c.}, \raisetag{18pt}
\end{split}
\end{equation}
where non-hermitean operators are in curly brackets.
We follow the conventions of~\cite[Eq.$(2.1)$]{Ferreira:2023dke} and the earlier literature throughout. 

In addition, we collect the Higgs fields and their complex conjugates in a single vector 
\begin{equation}
\vec{\Phi}:=\left(\Phi_1,\Phi_2,\Phi_1^*,\Phi_2^*\right)^\mathrm{T}\;.
\end{equation}  
This has the benefit that both, flavor- and CP-type transformations act linearly on $\vec\Phi$ and we can seamlessly combine them by matrix multiplication~\cite{Holthausen:2012dk}.\footnote{%
We drop space-time arguments throughout for notational convenience. For CP-type transformations it is understood that $(t,x)\mapsto(t,-x)$ in addition to the matrix action on $\vec\Phi$ (and the usual action of the complex conjugation outer automorphism on gauge fields and fermions~\cite{Grimus:1995zi,Trautner:2016ezn}).}
To address both scalar doublets simultaneously, we also define
\begin{equation}
 \Phi:=\left(\Phi_1,\Phi_2\right)^\mathrm{T}\,,\quad\text{and}\quad \Phi^*:=\left(\Phi_1^*,\Phi_2^*\right)^\mathrm{T}\,,\quad\text{i.e.}\quad \vec{\Phi}\equiv\left(\Phi,\Phi^*\right)^{\mathrm{T}}.
\end{equation} 
Explicitly, the well known global symmetries of the 2HDM~\cite{Ferreira:2009wh,Ferreira:2010yh,Branco:2011iw}
then act as 
\begin{align}\label{eq:RegularTrafos}
    \vec{\Phi}\mapsto\mat{  S\,&  \mathbf{0}\\   \mathbf{0}&  S^*\,}
    \vec{\Phi}\,,\qquad\text{or}\qquad   
    \vec{\Phi}\mapsto\mat{  \mathbf{0}&  X\\   X^*&  \mathbf{0}}
    \vec{\Phi}\,,
\end{align}
for flavor- or CP-type transformations, respectively.
$S$ and $X$ here are unitary {$2\times 2$} matrices whose explicit form depends on the class of symmetry considered and choice of basis. Classes of symmetries in the 2HDM are commonly denoted as $\Z{2}$, $\mathrm{U}(1)$, $\mathrm{SU}(2)$, CP1, CP2, and CP3. 
For convenience we have collected explicit expressions in App.~\ref{sec:appendix}, where our choices are consistent with~\cite{Ferreira:2023dke} and the previous literature. 
We refer to the above transformations as \textit{regular}, in contrast to the \textit{goofy} transformations to be discussed next.\footnote{%
The mathematically unavoidable binary distinction between  \textit{regular} and \textit{goofy} here is done in the commonly accepted denomination of symmetry-breaking stances in boardsports like skateboarding, snowboarding or surfing. This avoids the awkward and impossible binary distinguishability of ``goofy'' and ``normal'' in common sense, as somewhat provocatively has been chosen as the title of this paper.}

For later reference, we remark that that the class of regular transformations in Eq.~\eqref{eq:RegularTrafos} stems from all thinkable most general unitary transformations of the type
\begin{align}\label{eq:MostGeneralTrafo}
    \vec{\Phi}\mapsto\mat{  A&  C \\   D^*&  B^*\,}
    \vec{\Phi}\,,
\end{align}
where $A$, $B$, $C$, $D$ are general unitary $2\times2$ matrices (and conjugations in the definition have been chosen to comply with the rest of this work). The requirement of not mixing fields with different hypercharges selects the block diagonal ($C=D=0$) or anti-block diagonal ($A=B=0$) forms.
The additional requirement of exactly invariant kinetic terms imposes the conditions $B=A$ and $D=C$ on top~\cite{Doring:2024kdg}, ending up with Eqs.~\eqref{eq:RegularTrafos}. More general transformations allow for a non-trivial transformation of the kinetic terms, to be discussed~now.
\section{Goofy transformations}\label{sec:goofy}
Ferreira, Grzadkowski, Ogreid, and Osland~\cite{Ferreira:2023dke} observed that the 2HDM possesses RG stable points defined by the parameter relations
\begin{equation}\label{eq:goofy_relation}
 m_{11}^2=-m_{22}^2\,,\qquad \lambda_1=\lambda_2\,,\qquad\lambda_6=-\lambda_7\,.
\end{equation}
These relations do not correspond to any of the well-known regular symmetries. To nonetheless establish the above parameter relations from an active transformation, FGOO devised the GOOFy transformation
\begin{equation}\label{eq:goofy_trafo}
 \Phi_1\mapsto-\Phi_2^*\,,\quad
 \Phi_2\mapsto\Phi_1^*\,,\quad
 \Phi_1^*\mapsto\Phi_2\,,\quad
 \Phi_2^*\mapsto-\Phi_1\,.
\end{equation}
Enforcing this transformation as a symmetry on the 2HDM scalar potential reproduces~\eqref{eq:goofy_relation}.
However, on the canonical kinetic terms
\begin{equation}\label{eq:kinetic_1}
 \mathcal{L}_K=\left(D_\mu\Phi_1\right)^\dagger\!\left(D^\mu\Phi_1\right)+\left(D_\mu\Phi_2\right)^\dagger\!\left(D^\mu\Phi_2\right)\;,
\end{equation}
the transformation~\eqref{eq:goofy_trafo} acts as a global sign flip 
\begin{equation}
\mathcal{L}_K\mapsto-\mathcal{L}_K\;. 
\end{equation}
Hence, simply imposing it as a symmetry would imply $\mathcal{L}_K=0$, which is not a good premise for a dynamical theory. 
For that reason, FGOO devised to extend the goofy transformation to the space-time coordinate as $x_\mu\mapsto\I x_\mu$ and gauge fields $A_\mu\mapsto\pm\I A_\mu$~\cite{Ferreira:2023dke}, which would render the kinetic terms invariant and could lead to a consistent, yet somewhat goofy, transformation of the overall QFT. 

We recall that the original question is, \textit{why}~\eqref{eq:goofy_relation} corresponds to an RG stable point to all orders in the scalar couplings, and gauge couplings as we will see. The purpose of this note is to illuminate why that is the case independent of the eventual existence of exotic space-time symmetries.

To advance this, note that in the notation of the previous section, the goofy transformation~\eqref{eq:goofy_trafo} takes the form
\begin{align}\label{eq:goofy_trafo_2}
    \vec{\Phi}\mapsto\mat{  \mathbf{0}&  \varepsilon\\ -\varepsilon&  \mathbf{0}}
    \vec{\Phi}~\equiv~\mathcal{G}\,\vec{\Phi}\;,\qquad\text{where}\qquad
    \varepsilon:=\mat{0&-1\\1&0}\;,
\end{align}
and we have defined the $4\times4$ transformation matrix $\mathcal{G}$. This is a CP-type transformation of order $4$, but it does not fit the pattern of the regular transformations of Eq.~\eqref{eq:RegularTrafos}. The reason is simply that the regular transformations \eqref{eq:RegularTrafos} are constrained precisely in such a way as to leave invariant~$\mathcal{L}_K$.
\section{General unitary basis changes}
\label{sec:GUBC}
We want to explicitly show that there is nothing inherently ``wrong'' or unphysical about a theory with one or several ``wrong-signed'' kinetic terms. Consider unitary linear field redefinitions of the original theory starting from canonical kinetic terms.\footnote{%
We will discuss the most general case starting from a non-canonical basis for the kinetic terms in Sec.~\ref{sec:noncanonical}.}
In order not to mix up fields with different (hypercharge) gauge quantum numbers, we consider \textit{passive} field redefinitions of the form
\begin{align}\label{eq:basistrafo}
    \vec{\Phi}'=\mathcal{V}\,\vec{\Phi}=\mat{  V\,&  \mathbf{0}\\   \mathbf{0}&  U^*\,}
    \vec{\Phi}\,,
\end{align}
with unitary {$2\times 2$} matrices $V$ and $U$. 
The kinetic terms in the starting basis can be written as
\begin{equation}\label{eq:kinetic}
\mathcal{L}_K~=~\frac12\,\mat{\!D_\mu\Phi \\ \!\;D_\mu^*\Phi^*}^{\mathrm{T}}\mathcal{K}\;\mat{\!D_\mu\Phi \\ \!\;D_\mu^*\Phi^*}
\quad\text{with}\quad
\mathcal{K}:=\mat{\mathbf{0} & \mathbbm{1}\\ \mathbbm{1}&\mathbf{0}}\;.
\end{equation}
The gauge covariant derivatives are not decisive here and it suffices to discuss a term
\begin{equation}
 \widetilde{\mathcal{L}}_K~=~\frac12\,\vec{\Phi}^{\mathrm{T}}\,\mathcal{K}\,\vec{\Phi}\;,
\end{equation}
which contains the same $\mathcal{K}$ as the actual kinetic term, and also transforms in the same way under redefinitions of $\vec{\Phi}$.
After the general field redefinition~\eqref{eq:basistrafo}, $\widetilde{\mathcal{L}}_K$ reads
\begin{equation}\label{eq:dummy_K}
 \widetilde{\mathcal{L}}_K~=~\frac12\,\vec{\Phi}^{\prime\,\mathrm{T}}\,\mathcal{K}'\,\vec{\Phi}'\quad\text{with}\quad
 \mathcal{K}'=\mat{ \mathbf{0} & V^\mathrm{T}U^* \\ U^\dagger V & \mathbf{0}  }\;.
\end{equation}
The very same $\mathcal{K}'$ also appears in the actual kinetic term $\mathcal{L}_K$.
Consequently, the kinetic terms are invariant under a basis change if and only if $U=V$. Basis changes with $U=V$ are the ones which are usually
considered for switching between different bases of the explicit symmetry representation matrices, or to construct basis invariant physical quantities, see e.g.~\cite{Davidson:2005cw,Haber:2006ue,Haber:2010bw,Trautner:2018ipq,Bento:2020jei}.
By contrast, basis changes with $U\neq V$ will always change the form of the kinetic terms. Nonetheless, the physics described by a theory does not change under field redefinitions~\cite{Chisholm:1961tha,Kamefuchi:1961sb},\footnote{%
See also discussions in~\cite{Kallosh:1972ap, Passarino:2016saj, Manohar:2018aog,Criado:2018sdb} and references therein.}
especially when they act only linearly on the fields as considered here. This is not surprising as linear field redefinitions are just \textit{re-writings} of the original theory and no physical results of a theory can depend on the choice of conventions of formulation.

Hence, there is, in principle, nothing wrong with passive field redefinitions even if they make kinetic terms appear non-canonical. In other words, there is in principle nothing wrong about using a basis where an irreducible matrix representations of $\rep{r}$ and its conjugate (outer-automorphism related) irrep $\bar{\rep{r}}$ transform in mutually non-complex conjugate bases -- it is just usually a rather inconvenient choice (see e.g.\ the discussion in \cite[Sec.~3.2]{Chen:2014tpa} or \cite[App.~A.2]{Fallbacher:2015upf}). One of the main inconveniences is that propagators, in most generality, need to be treated in their matrix values forms which needs to be carried through all computations.

We stress that it is very important here to strictly separate passive from active transformations. We will exclusively focus on the situation that real fluctuations of $\phi_{1,\dots,8}$ (see Eq.~\ref{eq:realDOF}) maintain the action as a real functional. This means the real fields obey a Majorana-type constraint which in the initial basis reads $\phi_i=\phi_i^*$, or written in complex scalars: $\vec{\Phi}^*=\mathcal{K}\vec{\Phi}$. Under passive field-redefinitions  this constraint transforms covariantly, i.e.\ $\vec{\phi}'=\mathcal{V}_{8\times8}\vec{\phi}'^*$ and $\vec{\Phi}'^*=\mathcal{K}'\vec{\Phi}'$,
making sure that the action stays real. By contrast, for an active transformation, the hermiticity of the Lagrangian is generally lost if one allows for independent active transformations (mappings) of $\Phi$ and $\Phi^*$ while considering their unconstrained fluctuations.\footnote{In other words, one has to make sure that the number of independent real degrees of freedom is not changed in the mapping, which is not a danger for a $1:1$ rewriting of the theory in a passive transformation.}\enlargethispage{1cm}
This means that the active transformations are necessarily restricted by the constraint of maintaining a real valued action and we will come back to this in the discussion of Sec.~\ref{sec:generalgoofy}.

Let us now consider a basis change (a passive field redefinition) that exactly flips the sign of the kinetic terms, i.e.~$\mathcal{K}'=-\mathcal{K}$. This is the case if and only if $U=-V$. For example, consider the basis transformation with
\begin{equation}\label{eq:Sigma_3_BT}
 \vec{\Phi}'=\mat{  \sigma_3\,&  \mathbf{0}\\   \mathbf{0}&  -\sigma_3\,}\vec{\Phi}~\equiv~\Sigma_3\,\vec{\Phi}\,,
\end{equation}
where $\sigma_3=\mathrm{diag}(1,-1)$ is the third Pauli matrix. In the primed basis, the original goofy transformation \eqref{eq:goofy_trafo_2} takes the form
\begin{equation}
\vec{\Phi}'\mapsto \Sigma_3\mathcal{G}\Sigma_3^{-1}\vec{\Phi}'\equiv \mathcal{G}'\vec{\Phi}'.
\end{equation}
This is, of course, still a goofy transformation irrespectively of the sign-flip of the kinetic term under the basis change. The initial sign of the kinetic term has nothing to do with its transformation behavior under the goofy transformation.

\enlargethispage{1cm}
The couplings in the primed basis are related to the couplings in the starting basis as
\begin{equation}
 m_{11}^{\prime\,2}=-m_{11}^2\,,\qquad
 m_{22}^{\prime\,2}=-m_{22}^2\,,\qquad
 \lambda_6'=-\lambda_6\,,\qquad
 \lambda_7'=-\lambda_7\,,
\end{equation}
while all other couplings stay the same. 
Note that the goofy parameter relations Eq.~\eqref{eq:goofy_relation} stay invariant under the basis transformation with $\Sigma_3$.\footnote{%
Note that~\cite{Ferreira:2023dke} emphasized the basis invariance of~\eqref{eq:goofy_relation}, which is easy to understand in our notation from the fact that $\epsilon$ in Eq.~\eqref{eq:goofy_trafo_2} is an invariant tensor of \SU{2}. However, this only shows the invariance under \textit{regular} basis changes while $\mathcal{G}$ is generally not invariant under arbitrary \textit{goofy} basis changes.}
This is a manifestation of the fact that $\Sigma_3\mathcal{G}=\mathcal{G}\Sigma_3$, i.e.\ this specific field redefinition does commute with the goofy transformation.\footnote{%
Commutation up to a global hypercharge transformation or any other already present symmetry would be acceptable here. 
} 
This shows that we can flip the sign of the kinetic term by a field redefinition, here Eq.~\eqref{eq:Sigma_3_BT}, \textit{without altering} the goofy parameter relations. 
Hence, even though the goofy transformation flips the sign of the kinetic term, the sign of the kinetic term is not determined by imposing the goofy parameter relations.

Yet another way to phrase this is to say that the sign of the kinetic term is physical \textit{only} with respect to the signs of $m_{11}$, $m_{22}$, $\lambda_6$ and $\lambda_7$.\footnote{%
Alternatively, one may perform the sign-flipping basis change with $\mathcal{P}_{\mathcal{G}}$ (see Sec.~\ref{sec:generalgoofy} below) to conclude that the sign of $\mathcal{K}$ is only physical w.r.t.\ the signs of $m_{11}$, $m_{22}$, and $m_{12}$. This is in consistency with the fact that the sign of $m_{12}$ can be traded for signs of $\lambda_6$ and $\lambda_7$ by a regular basis transformation.}
This establishes that a theory which obeys the goofy parameter relations of Eq.~\eqref{eq:goofy_relation} is mathematically healthy, irrespective of the sign of the kinetic term. If one is unhappy about the sign, one can simply use a field redefinition to flip it, which transforms the other couplings but does not alter the goofy relations.  

It is important to establish the existence of a field redefinition which takes one back to a theory with canonical kinetic terms, because all presently available higher-loop derivations of RG equations start from canonical kinetic terms. Ensuring the existence of a canonical basis allows us to use the RGEs \textit{in that basis} without having to worry about subtleties that arise if one would have to absorb the goofy sign of a kinetic term in the wave function renormalization. If one would start with non-canonical kinetic terms, one would have to specify how the unusual sign is absorbed in wave function renormalization coefficients and we will come back to this discussion in Sec.~\ref{sec:noncanonical}.

\section{Global-sign-flipping goofy transformations}\label{sec:generalgoofy}
We extend the discussion of the previous section from passive to active transformations. For this, consider the most general possible unitary transformation, Eq.~\eqref{eq:MostGeneralTrafo}, 
again imposing homogeneous transformation in hypercharge but dropping the requirement of invariant kinetic terms.
Such transformations act on the fields as 
\begin{align}\label{eq:GoofyTrafos}
    \vec{\Phi}\mapsto\mat{  A\,&  \mathbf{0}\\   \mathbf{0}&  B^*\,}
    \vec{\Phi}\,,\qquad\text{or}\qquad   
    \vec{\Phi}\mapsto\mat{  \mathbf{0}&  C\\   D^*&  \mathbf{0}}
    \vec{\Phi}\,.
\end{align}
Just as for the regular case, also the goofy transformations either act as a flavor-type or CP-type transformation. They respectively act on the canonical kinetic terms as
\begin{equation}
 \mathcal{K}\mapsto\mathcal{K}'=\mat{ \mathbf{0} & A^\mathrm{T}B^* \\ B^\dagger A & \mathbf{0}  }\,,\qquad\text{or}\qquad
 \mathcal{K}'=\mat{ \mathbf{0} & D^\dagger C \\ C^\mathrm{T}D^* & \mathbf{0}  }\,.
\end{equation}
This can produce more than just a global sign flip of $\mathcal{K}$. We will return to the most general case below but focus for the rest of this section on purely global-sign-flipping goofy transformations, implying that $\mathcal{K}'=-\mathcal{K}$, i.e.\ $B=-A$ and $D=-C$. 

Just like for the regular 2HDM symmetries, it is possible to classify all of the goofy transformations but this is not the goal of this work.\footnote{%
Just as for regular transformations~\cite{Battye:2011jj, Pilaftsis:2011ed}, one may also consider goofy transformations that do not commute with hypercharge or other gauge transformations. Knowledge about such transformations (like e.g.\ the well-known SM custodial symmetry) is of phenomenological interest but beyond the scope of this work.} As a start, we can take the goofy versions of any of the well-known classes of 2HDM symmetries. This means we can use for $A$ or $C$ the same matrices as for the well-known regular symmetry transformations, but take the opposite-than-normal sign for the second block ($B$ or $D$, respectively) to render the transformation goofy. This modifies the regular transformations of Eq.~\eqref{eq:RegularTrafos} and gives rise to their goofy versions,
\begin{align}\label{eq:GoofyTrafos2HDM}
    \vec{\Phi}\mapsto\mat{  S\,&  \mathbf{0}\\   \mathbf{0}&  -S^*\,}
    \vec{\Phi}\,,\qquad\text{or}\qquad   
    \vec{\Phi}\mapsto\mat{  \mathbf{0}&  X\\   -X^*&  \mathbf{0}}
    \vec{\Phi}\,,
\end{align}
with possible matrix choices listed in App.~\ref{sec:appendix}. 
The corresponding parameter relations of these transformations called $\mathrm{CP1}_G$, $\mathbbm{Z}_{2,G}$ and so on in the obvious way, are collected in Tab.~\ref{tab:Goofy_relations}. 

The relations shown in the upper part of Tab.~\ref{tab:Goofy_relations} are RG stable to all orders under scalar and gauge corrections, even though they do not correspond to any previously known symmetries of the 2HDM (some of them do accidentally imply regular symmetries as can straightforwardly be checked by their necessary and sufficient basis invariant conditions~\cite{Bento:2020jei}). This all-order stability can be understood by our arguments laid out in Sec.~\ref{sec:RGstability}. 

We have explicitly confirmed the RG stability of the relations in the $\overline{\mathrm{MS}}$~\cite{Bardeen:1978yd} renormalization scheme at the three-loop order in scalar corrections using~\cite{Bednyakov:2018cmx}, and three-loop order in gauge corrections using RGBeta~\cite{Thomsen:2021ncy,*Poole:2019kcm} (based on \cite{Pickering:2001aq,*Bednyakov:2021qxa,*Davies:2021mnc,*Steudtner:2024teg} and \cite{Schienbein:2018fsw}). Corrections from Yukawa couplings depend on whether or not the goofy transformations are imposed on them.
Investigating the Yukawa corrections in detail is of great phenomenological interest -- especially also regarding the vacuum structures of the goofy symmetric theories -- but this beyond the scope of this work.

The ``original'' goofy transformation $\mathcal{G}$, as discovered by~\cite{Ferreira:2023dke}, is included here as ``a goofy CP2'' or simply ``$\mathrm{CP2}_G$'' based on the identical matrix generators in block~$C$. All the other transformations are new goofy transformations of the 2HDM.
We stress that the transformations in Tab.~\ref{tab:Goofy_relations} are \textit{genuinely new} and not the same as the transformations called ``0CP1'', ``0\Z{2}'' \textit{etc.} in~\cite{Ferreira:2023dke}, which arise as combination of $\mathcal{G}$ (i.e.~$\mathrm{CP2}_G$) and all possible kinds of regular-type transformations. Just like $\Sigma_3$ is not based on a product of the original goofy and a non-trivial regular transformation, none of the transformations in Tab.~\ref{tab:Goofy_relations} is. This is not in contradiction to the fact that the corresponding transformation matrices can be factored as product of a goofy and a regular transformation matrix. 
\begin{table}[t]
\begin{center}
\scalebox{0.75}{
\hspace{-0.85cm}
\begin{tabular}{lcccccccccccl}
\toprule[1pt]
\centering
\multirow{2}{*}{\shortstack{Goofy \\ trafo.}} & \multicolumn{10}{c}{parameter relations} &
\multirow{2}{*}{\shortstack{accidental \\ regular sym.}} & \multirow{2}{*}{\shortstack{leading vanishing \\ co-/in-variants}} \\ 
& $m_{11}^2$ & $m_{22}^2$ & $m_{12}^2$ & $\lambda_{1}$ & $\lambda_{2}$ & $\lambda_{3}$ & $\lambda_{4}$ & $\lambda_{5}$ & $\lambda_{6}$ & $\lambda_{7}$ & & \\
\hline
$\mathcal{P}_G$ & $0$ & $0$ & $0$ & & & & & & & & $-$ & $Y=0$ \\
$\mathbbm{Z}_{2,G}\equiv\Sigma_3$ & $0$ & $0$ & & & & & & & $0$ & $0$ & $-$ & $YT=0$, $QYT=0$ \\
$\mathrm{CP1}_{G}$ & $0$ & $0$ & $-{m_{12}^{2}}^*$ & & & & & $\lambda_5^*$ & $\lambda_6^*$ & $\lambda_7^*$ & $-$ & $YT=0$, $QYT=0$ \\
$\mathrm{CP2}_{G}\equiv\mathcal{G}$ & & $-m_{11}^2$ & & & $\lambda_1$ & & & & & $-\lambda_6$ & $-$ & $T=0$ \\
$\U1_{G}$ & $0$ & $0$ & $0$ & & & & & $0$ & $0$ & $0$ & $\U1$ & $Y=0$ \\
$\mathrm{CP3}_{G}$ & $0$ & $0$ & $0$ & & $\lambda_1$ & & & $\lambda_1-\lambda_3-\lambda_4$ & $0$ & $0$ & CP2 & $Y=T=0$ \\
$\SU2_{G}$ & $0$ & $0$ & $0$ & & $\lambda_1$ & & $\lambda_1-\lambda_3$ & $0$ & $0$ & $0$ & $\SU2$ & $Q=Y=T=0$ \\[0.1cm]
\hline\\[-0.4cm]
$\mathrm{CP2}_{G}^{\mathrm{soft}}$ & & & & & $\lambda_1$ & & & & & $-\lambda_6$ & $-$ & $T=0$
\\[0.1cm]
\hline\\[-0.4cm]
$\Z{2,G}^{-}$ & & $\phantom{{}^{\dagger2}}0^{\dagger2}$ & $0$ & & & $\phantom{{}^{\dagger1}}0^{\dagger1}$ & $\phantom{{}^{\dagger1}}0^{\dagger1}$ & & $0$ & $0$ & $\Z2$ & $-$ \\
$\mathrm{CP4}_G$ & & $\phantom{{}^{\dagger2}}0^{\dagger2}$ & $0$ & & & $\phantom{{}^{\dagger1}}0^{\dagger1}$ & $\phantom{{}^{\dagger1}}0^{\dagger1}$ & $\lambda_5^*$ & $0$ & $0$ & $\Z2$ & $-$ \\
$\Z{4,G}^{-}$ & & $\phantom{{}^{\dagger2,g}}0^{\dagger2,g}$ & 0 & & & $\phantom{{}^{\dagger1,g}}0^{\dagger1,g}$ & $\phantom{{}^{\dagger1,g}}0^{\dagger1,g}$ & $0$ & $0$ & $0$ & $\U1$ & $-$ \\
\bottomrule[1pt]
\end{tabular}}
\end{center}
\caption{\label{tab:Goofy_relations}
Goofy transformations and their corresponding parameter constraints, accidental regular symmetries, and vanishing leading-order basis co- or invariants. Relations in the top part are RG stable to all orders in scalar and gauge quantum  corrections. Relations in the bottom part are not stable to all order and receive radiative corrections in couplings marked by a dagger ($\dagger$), and a number that indicates the loop order of first corrections ($g$ means that only gauge corrections arise).}
\end{table}
We demonstrate this by the following example. Consider the goofy transformation 
\begin{align}\label{eq:minimal_GoofyTrafo}
    \vec{\Phi}\mapsto\mat{  \mathbbm{1}\,&  \mathbf{0}\\   \mathbf{0}&  -\mathbbm{1}\,}
    \vec{\Phi} ~\equiv~\mathcal{P}_G\,\vec{\Phi}\,.
\end{align}
One may refer to this as a ``goofy parity'' $\mathcal{P}_G$ (no space-time transformations are implied), and we list the corresponding parameter relations in Tab.~\ref{tab:Goofy_relations}. Multiplying this transformation with any of the regular flavor- or CP-type symmetries turns these into their goofy versions. Nonetheless, separately imposing $\mathcal{P}_G$ and the corresponding regular transformation on the same potential implies more restrictive conditions on the parameters than just imposing the corresponding goofy symmetry that arises as their product. Take, for example, the well-known regular flavor-type $\mathbbm{Z}_2$ symmetry (which is generated by a $A=B=\sigma_3$, see App.~\ref{sec:appendix}, and requires parameter relations $m_{12}=0$, $\lambda_6=0$, $\lambda_7=0$). All of the transformations, $\mathcal{P}_G$, $\Sigma_3$ and $\mathbbm{Z}_2$, commute, and for the corresponding matrix generators one finds relations like $\mathcal{P}_G\times\mathbbm{Z}_2=\Sigma_3$. Yet, imposing $\mathcal{P}_G$ and imposing $\mathbbm{Z}_2$ separately is not the same as imposing $\Sigma_3$ but leads to more restrictive relations.

Limiting oneself to goofy transformations with a kinetic-term sign flip, one may also notice that a combination of two goofy transformations (or the even-times repeated action of a goofy transformation) is always a regular transformation, hence, by itself obviously also poses different (typically less restrictive) requirements as combining the individual requirements of both goofy transformations. In this sense, goofy transformation matrices may be viewed as ``roots'' of regular transformation matrices (not implying that their parameter relations are less restrictive).

\enlargethispage{1cm}
Finally, we note that goofy transformations, just like regular transformations, imply relations on the basis invariants of the theory that are in one-to-one correspondence with the respective symmetry and hold to all orders in the RG evolution. We refer to the basis invariants of the 2HDM by using their basis covariant constituents, namely the triplets  $Y$, $T$ and the quintuplet $Q$, following the conventions of~\cite{Trautner:2018ipq,Bento:2020jei} (see e.g.~\cite{Bednyakov:2025sri} for the translation to different notations). Just like CP-type regular transformations, goofy transformations are not subgroups of the regular flavor-type basis transformations that have been used to construct the invariants. Hence, basis invariants can transform non-trivially under goofy transformations (just like for regular CP-type transformations). In the last column of Tab.~\ref{tab:Goofy_relations} we collect some remarkable co- or invariant relations implied by the respective goofy transformation (the listed conditions are generally not $1:1$ and additional relations of invariants are implied which are not shown).

\section{General sign-flipping goofy transformations}\label{sec:goofybeyondsigns}
We now extend the discussion beyond global sign flips
of the kinetic term and consider the most general possible goofy transformations that commute with hypercharge, given by Eq.~\eqref{eq:GoofyTrafos} with general unitary matrices $A$, $B$, $C$, and $D$. This also serves to set the stage to discuss conditions for the all-order RG stability in the next section. 

In order to keep the Lagrangian density real under a general active flavor- or CP-type transformation, the transformation matrices are restricted to fulfill that $A^{\mathrm{T}}B^*$ or $D^\dagger C$, respectively, is hermitean.\footnote{%
In the future, it would also be interesting to study goofy transformations in the context of non-hermitean Hamiltonians with nonetheless real eigenvalues, such as PT-symmetric theories see e.g.~\cite{Bender:2007nj}.}
Being hermitean and unitary, these matrices can only have eigenvalues $\kappa_{1,2}=\pm1$. These eigenvalues can be exposed by a \textit{regular} basis change, i.e.\ we can, in principle, always work in a basis where the effect of the most general goofy transformation on the canonical kinetic terms is given by
\begin{equation}\label{eq:LKtrafo}
 \mathcal{L}_K~\mapsto~
 \kappa_1\left(D_\mu\Phi_1\right)^\dagger\!\left(D^\mu\Phi_1\right)+
 \kappa_2\left(D_\mu\Phi_2\right)^\dagger\!\left(D^\mu\Phi_2\right)\,
 \qquad\text{with}\qquad\kappa_1=\pm1,\;\kappa_2=\pm1\;.
\end{equation}
The case $\kappa_1=\kappa_2=1$ corresponds to regular transformations. Examples for the the case with $\kappa_1=\kappa_2=-1$ that we refer to as global-sign-flipping goofy transformations were treated in the previous section. The most curious case is the one with relative sign flips of the kinetic terms, here corresponding to the situation with $\kappa_1=-\kappa_2=\pm1$. As examples for
goofy transformations with relative sign flip of the kinetic terms we introduce
\begin{align}
 &\Z{2,G}^{-}:& &A=\mat{1 & 0\\ 0 & -1},~B^*=\mathbbm{1},& &C=D^*=0\;,& \\
 &\mathrm{CP4}_G^{-}:& &A=B^*=0,& &C=\mat{1 & 0 \\ 0 & -1},~D^*=\mathbbm{1}\;.& \\
 &\Z{4,G}^{-}:& &A=B^*=\mat{1 & 0\\ 0 & i},& &C=D^*=0\;.&
\end{align}
The corresponding parameter relations are summarized in the lower part of Tab.~\ref{tab:Goofy_relations}.

The sign flip of one or several of the gauge-kinetic terms means that goofy transformations are generally broken by the propagators, and by non-zero gauge couplings. 
Such a breaking can be ``soft'', in the sense that it does
not necessarily feed back to other operators in the RG flow. In situations where this is the case, goofy parameter relations implied on the potential of the theory can be stable to all orders. 

\enlargethispage{1cm}
In the next section, we will discuss examples for both cases, namely goofy parameter relations that are exact to all orders in gauge- and scalar corrections, and examples where the explicitly goofy-breaking gauge-kinetic terms feed back to re-generate other goofy violating operators. 
We will see that global-sign-flipping goofy transformations 
lead to parameter relations that are stable to all orders, while parameter relations derived from goofy transformations that lead to a relative sign flip of one or several kinetic terms generally receive radiative corrections and, hence, are not RG stable.

\section{RG stability of regular and goofy transformations}
\label{sec:RGstability}
We now return to the original question, which was why the goofy parameter relations are RG stable, even though they do not correspond to symmetry transformations in the regular sense. To understand this, it is helpful to understand \textit{why}, in the first place, even regular symmetry transformations do imply RG stable parameter relations. Our best argument is an extension of the original technical naturalness argument brought forward by 't\,Hooft~\cite{tHooft:1979rat}. It is a folk wisdom that the beta function of a symmetry-breaking parameter has to be proportional to the symmetry-breaking parameter itself, such that if the breaking is set to zero also the corresponding beta function vanishes. A formal generalization of this argument based on general grounds will be presented in~\cite{deBoer:2025xx} (this is a symmetry-based, non-perturbative argument that can be applied to any QFT). Here, we will only sketch the argument but already use it to establish the all-order stability of some of the identified goofy transformations. In other cases, our argument will allow us to identify why and how goofy relations are violated in the RG flow, and at what order corrections have to be expected. 

\enlargethispage{1cm}
Consider a possible symmetry transformation (including also goofy transformations) that acts on an un-symmetric\footnote{%
By ``un-symmetric'' we mean a theory that has fields in multiplets unconstrained by any symmetry, like the two Higgs doublets in the most general 2HDM, or the three copies of SM generations in flavor space.} theory. The prospective symmetry transformation in most cases acts as an outer automorphism\footnote{%
For a brief introduction to outer automorphisms see~\cite[2.2]{Doring:2024kdg} and references therein, which besides extensions by outer automorphisms also discusses the second (exhaustive) possibility of so-called ``unorthodox'' extensions. We limit ourselves to outer automorphisms here, noting that the only thing one needs to understand about them for the present discussion is that they act as a linear map within the parameter space of a theory.} on the un-symmetric theory. This implies that the transformation acts as a linear transformation in the parameter space of the theory~\cite{Fallbacher:2015rea} (for goofy transformations, including the action on the most general kinetic term). In particular, this means that the parameters of the un-symmetric theory transform covariantly under the action of any perhaps-to-be-imposed-as-a-symmetry transformation.\footnote{%
If there is no previously existing symmetry in the theory, say starting from the most general unconstrained 2HDM scalar potential, the options for outer automorphisms include all possible basis changes of the model (including goofy basis changes). If there is a previously existing symmetry (like, strictly speaking, the global $\U{1}$ hypercharge here), the possible transformations are limited to outer automorphisms of that symmetry.} This covariant transformation behavior is reflected as a symmetry of the full coupled system of beta functions. 

Consider the full coupled system of beta functions ($\lambda_i$'s here should be thought of running over all couplings and masses), 
\begin{equation}\label{eq:beta}
\beta_{\lambda_i}~\equiv~\mu\frac{\mathrm{d}\,\lambda_i}{\mathrm{d} \mu}~=~f_{i}(\lambda_1,\lambda_2,\dots)\;.
\end{equation}
\textit{By definition}, the beta functions on the l.h.s.\ are linear in the couplings (differentiation is a linear operator).
Hence, under outer automorphisms, the beta functions have to transform covariantly and in the same representation as the couplings themselves. This means that the non-linear expressions $f_{i}(\lambda_1,\lambda_2,\dots)$ on the r.h.s.\ of Eq.~\eqref{eq:beta} \textit{have to} transform covariantly under the outer automorphism.

This gives rise to strong, all order exact constraints on the beta functions. The more independent outer automorphisms are possible, the stronger the overall constraints on the beta functions. Very often this fully determines the complete functional form of the beta functions in terms of non-trivially transforming covariant combination of couplings.  For simple algebraic reasons, the number of possible covariant structures in the beta functions terminates at typically rather low orders such that the structure of the beta function can fully be determined at the non-perturbative level.

If the corresponding outer automorphism transformation were to be \textit{enforced} as a symmetry, this corresponds to \textit{requiring} that all non-trivially transforming covariant combinations of couplings vanish. This ensures the absence of symmetry-breaking couplings and, by above argument, also the vanishing of all beta functions that correspond to non-trivially transforming combinations of couplings, very much in 't\,Hooft's sense, but never spelled out as explicitly.

\enlargethispage{1cm}
Note that our statements on the decomposition of the beta functions neither require that any transformation is imposed, nor that we are perturbatively close to the limit where any of them would become a symmetry; the system of beta functions has to transform covariantly under \textit{all possible} outer automorphisms in any case, even if we are parametrically far away from points that realize the symmetry. 

As simple concrete example in the 2HDM, consider the covariant combination of quartic couplings that transforms as a triplet under the most general basis changes, commonly denoted as $\vec\Lambda$.\footnote{%
$\vec\Lambda\propto\left\{\mathrm{Re} (\lambda_6+\lambda_7), -\mathrm{Im}(\lambda_6+\lambda_7),\frac{1}{2}(\lambda_1-\lambda_2)\right\}^\mathrm{T}$, see e.g.~\cite{Bednyakov:2018cmx} and references therein. $\vec\Lambda$ corresponds to $T$ in the invariant language used above and in~\cite{Trautner:2018ipq,Bento:2020jei}.} Our arguments imply that 
\begin{equation}\label{eq:RGLambda}
 \beta_{\vec{\Lambda}}~\propto~\vec\Lambda\;,
\end{equation}
and this is a non-perturbative, all-order statement. The outer automorphism that matters to derive this statement is the most general regular basis transformation. In principle, $\beta_{\vec{\Lambda}}$ could have contributions from any combination of couplings that transform like a triplet, such as $\vec{M}$ (corresponding to $Y$) or triplets appearing in the contractions of five-plet covariant $\tilde\Lambda$ (corresponding to $Q$) with itself, i.e.\ $\rep{3}\subset\left(\rep{5}\otimes\rep{5}\otimes\dots\right)$. However, $\vec{M}$ cannot enter the beta function of $\vec\Lambda$ for dimensional reasons (which, in fact, is our very argument applied to the scaling outer automorphism contained in the conformal group), while self-contractions of $\tilde{\Lambda}$ do not enter for algebraic reasons: all triplet contractions of an individual fiveplet with itself vanish.  For completeness, we remark that the proportionality factor in the above relation is generally matrix-valued and can take coefficients in the space of the unit matrix, $\tilde\Lambda$, and $\tilde\Lambda^2$, see~\cite{Bednyakov:2018cmx}, and can itself further be constrained by using other outer automorphisms.

The proportionality in \eqref{eq:RGLambda} shows that $\vec\Lambda=0$ is a fixed point of the RG flow, irrespective of any other assumption. 
This fixed point can be enforced by any transformation that maps $\vec\Lambda\mapsto-\vec\Lambda$, corresponding to $\lambda_1=\lambda_2$ and $\lambda_6=-\lambda_7$.  All known transformation that require these parameter relations, for example CP2 or $\mathrm{CP2}_G$, do require relations on the mass terms in addition. However, none of the additional relations is required to make \eqref{eq:RGLambda} vanish.
This implies that even in the case that $\mathrm{CP2}$ or $\mathrm{CP2}_G$ is explicitly broken by mass terms, see entry $\mathrm{CP2}_G^{\mathrm{soft}}$ in Tab.~\ref{tab:Goofy_relations}, the statement of $\vec\Lambda=0$ as an RG fixed point holds to all orders in scalar and gauge quantum corrections.

Let us now turn to goofy transformations. Our argument of covariant transformation of couplings, masses and beta functions above can, in exactly the same manner, be applied to the ``couplings'' of the gauge-kinetic terms, i.e.\ the hermitean matrix of wave function renormalization coefficients. As the kinetic terms are left invariant by regular transformations in any basis, tracking the appearance of wave function renormalization coefficients in the RGs (in a closed form) is usually not considered to be very important. But this is important for goofy transformations (and also field rescaling transformations), as these have a non-trivial action on the kinetic terms. If we do not allow the kinetic terms to vanish (as we will assume everywhere here) we are unavoidably in a regime where the sign-flipping goofy transformations are explicitly broken. This does not mean that their corresponding parameter relations cannot be stable under RG running, as above example for the case of an explicitly, softly broken regular transformation suggests, and as we will demonstrate now. 

In principle, we would like to track the transformation behavior of all couplings, including wave function renormalization coefficients, under goofy transformations in the most general basis. But this is currently not possible as all available higher-loop derivations of RG equations (RGEs) start from canonical kinetic terms. Hence, we also have to adopt the canonical basis here and defer the most general case to the next section.

Working in the canonical basis means that goofy transformations impose a sign flip of one or several of the gauge-kinetic terms, corresponding to signs of $\kappa_i$ in Eq.~\eqref{eq:LKtrafo}. The sign flip in the full gauge-kinetic term corresponds to a sign flip in the respective propagator and gauge coupling vertices (linear and quadratic). Hence, in the canonical basis, we can manually track the effect of goofy transformations on the RG beta functions by a simple propagator and gauge vertex counting in the Feynman diagrams of the quantum corrections.

Imagine a generic linear combination of couplings $\lambda_{-}$ that transforms with a sign under a generic goofy transformation in the canonical basis.\footnote{%
Covariant transformations in higher-dimensional irreps are generally possible. Since we want to track the transformation of the gauge-kinetic terms in the canonical basis here, we focus on sign-flipping transformations.}
By definition, the goofy transformation will also flip one (or several) signs of the 
gauge-kinetic terms, i.e.\ signs of propagator and linear and quadratic gauge coupling vertices. Hence, by our arguments above, the beta function of $\lambda_{-}$ can only be a function that itself transforms in the same way as $\lambda_{-}$, i.e.\ schematically,   
\begin{equation}\label{eq:Goofy_beta}
 \beta_{\lambda_{-}}~ \propto ~ \lambda_{-}\,f_{+}(\lambda_i) + 
 \tilde{\lambda}_{-}\,g_{+}(\lambda_i) + \kappa_i^{\mathrm{odd}}\,h_{+}(\lambda_i)\;.
\end{equation}
Here, $f_+$, $g_+$ and $h_+$ are arbitrary functions which are invariant under the goofy transformation, $\tilde{\lambda}_{-}$ denotes a generic combination of other couplings that transforms odd under the goofy transformation,\footnote{%
For the original goofy $\mathrm{CP2}_G$ transformation concrete examples would be $\lambda_{-}=\lambda_1-\lambda_2$ and $\tilde\lambda_{-}=\lambda_6+\lambda_7$.} and  $\kappa_i^{\mathrm{odd}}$ corresponds to terms that transform with a sign under the goofy transformations because they originate from diagrams with an in-total odd number of sign-flipping propagators and gauge coupling vertices. Imposing goofy relations on all parameters of a theory (besides the gauge-kinetic terms) implies we sit at points where $\lambda_-,\tilde\lambda_-=0$. Hence, whether or not the goofy parameter relations are RG stable depends on how the sign-flipping, explicitly goofy breaking kinetic terms and gauge vertices enter the beta functions of goofy breaking parameters $\lambda_{-}$, i.e.\ especially on the form of function $h_{+}$.

Let's discuss the situation in the 2HDM. First we focus on goofy transformations that lead to a \textit{global} sign flip of the full gauge-kinetic terms. That is, a simultaneous sign flip of both $H_1$ and $H_2$ propagators and global sign flip of all (linear and quadratic) gauge vertices. It turns out that for the 2HDM (which, crucially, does not have scalar trilinear couplings) the \textit{total} number of propagators and gauge coupling vertices (linear and quadratic) entering the quantum corrections to \textit{any} coupling is always even. This means that the negative sign under goofy transformations which is required to make a contribution to $\beta_{\lambda_{-}}$ can neither be carried by a propagator nor by a gauge coupling vertex because the total number of signs arising from them come in even powers. Hence, it can only stem from goofy violating couplings $\lambda_{-}$ or $\tilde\lambda_{-}$, implying that $h_{+}=0$ in all beta functions.
In other words, in the 2HDM \textit{the global sign of the gauge-kinetic term does not enter the beta functions}.\footnote{%
I am grateful to Apostolos Pilaftsis for confirming this observation.} Consequently, if we require the (global-sign-flipping) goofy transformation to be conserved by all but the gauge-kinetic couplings, this is enough to make the beta functions of the goofy-breaking couplings vanish to all order. The goofy-violating gauge and kinetic terms cannot regenerate other goofy-violating couplings, simply because they do not enter the beta functions in the required way (namely, if they enter, they enter only proportional to other goofy violating couplings). This explains why the parameter relations implied by the global-sign-flipping goofy transformations are stable to all orders, even though the symmetry is \textit{explicitly} broken by the gauge-kinetic terms. In other words, the gauge-kinetic terms only \textit{softly} break the goofy symmetry and this soft symmetry breaking does not regenerate other, ``harder'' symmetry-breaking parameter combinations in the RG flow. This is similar\footnote{%
A slight difference is that the soft-breaking operator in the goofy case is the gauge-kinetic $d=4$ naively marginal operator, while in the regular symmetry example the soft-breaking operator is a $d=2$ relevant operator.} to our example above with softly broken CP2 regular symmetry, yet exactly vanishing $\beta_{\vec\Lambda}=0$ (if $\vec{\Lambda}=0$) to all orders.

Now consider the case of a goofy transformation that induces a \textit{relative} sign flip of the gauge-kinetic terms of $H_1$ and $H_2$. In this case, only one of the kinetic terms and gauge couplings to one of the Higgses violate the goofy transformation explicitly. This means our simple counting above does not hold, as one has to count the total number of gauge vertices and propagators for each of the fields separately. Even if the total number of propagators and gauge vertices is always even in the 2HDM (as used in the previous paragraph), each field may contribute with an odd number of propagators and/or odd number of gauge vertices as long as this is compensated by an odd number of propagators and vertices of the other field. In other words, \textit{the relative sign of the gauge-kinetic terms} does \textit{enter the beta functions}. In this case, it is possible to have a contribution to the beta functions $h_{+}\neq0$
(and without a proportionality in $h_{+}$ to $\lambda_{-}$ or $\tilde\lambda_{-}$).  Such a contribution will not vanish even if goofy parameter relations are imposed on all but the kinetic term. Hence, goofy transformations with a relative sign flip give rise to parameter relations that are not all stable under RG running. We have explicitly confirmed this for the relative-sign-flipping goofy transformations introduced in the previous section. In Tab.~\ref{tab:Goofy_relations} we 
mark couplings that vanish by goofy transformations at tree level but get restored by quantum corrections with a dagger ($\dagger$) and a number that corresponds to the loop-order of their restoration (and a letter $g$ if the corrections are purely due to gauge couplings). We remark that not all goofy-breaking couplings get radiatively reintroduced, a fact that can be understood by the remaining accidental regular symmetries of these points. 

\enlargethispage{1cm}
This concludes our argument. It explains why global-sign-flipping goofy transformations in the 2HDM are stable to all order in RG flow, and it explains why goofy transformations with a relative sign flip of the kinetic terms in the 2HDM generally receive radiative corrections and at which order.\footnote{%
We stress that~\cite{Ferreira:2023dke} already established the all-order RG stability (in scalar and gauge corrections) of relations~\eqref{eq:goofy_relation} based on the explicit (all-order) functional form of the RGEs derived in~\cite{Bednyakov:2018cmx}.
Our argument reproduces this conclusion but it is more general than that. In particular, using the outer automorphism nature of transformations constrains the beta functions without requiring their prior knowledge, and this generally applies to any model~\cite{deBoer:2025xx}.}
A completely analogous argument can be
made for the two-real-scalar-field toy model discussed in~\cite{Grzadkowski:2024,Haber:2025cbb} to show that the global-sign-flipping goofy transformation $\phi_1\mapsto i\phi_2$, $\phi_2\mapsto -i\phi_1$ there leads to relations that are RG stable to all orders.

We remark that the relative-sign-flipping goofy transformations uncover a very interesting special class of
quantum corrections that, here, are sensitive to the \textit{relative sign} of the gauge coupling vertices of different ``generations'' of Higgs flavor. In more general settings with fermions, which only have linear gauge interactions, the signs of fermion gauge coupling vertices correspond to the sign of the gauge coupling. We think the sensitivity of the relative signs of such terms is remarkable, as usually nothing is expected to be dependent on the sign of gauge couplings, which is often claimed to be unphysical. We have focused our discussion here exclusively to sign flips of the gauge-kinetic terms, as this is the case one is limited to if RGs are derived in the canonical basis. We briefly discuss more general cases now.

\section{RG running beyond the canonical basis}
\label{sec:noncanonical}
It is always possible to start from the canonical basis in order to compute quantum corrections and higher-loop RG running.\footnote{%
Strictly speaking, we will see that this is the case only for parameter points which do not sit precisely at a point which realizes a goofy symmetry that enforces a zero eigenvalue of the kinetic term.} However, the canonical basis may not be optimal to fully track the implications of a specific goofy transformation on the beta functions to all orders.\footnote{%
Also for other reasons the canonical basis may not be optimal, for example, if there are other symmetries at play (here in Higgs-flavor space), and the required basis change may bring them to an undesirable form, see also our discussion in Sec.~\ref{sec:GUBC} and refs.~\cite[Sec.~3.2]{Chen:2014tpa} or \cite[App.~A.2]{Fallbacher:2015upf}.} The superior way to properly track the non-perturbative constraints put by goofy transformations (and also scale transformations) on the RG flow, is to use the most general kinetic basis as a starting point.

Non-canonical kinetic terms in the 2HDM were previously considered in~\cite{Pilaftsis:1997dr,Ginzburg:2004vp,Ivanov:2006yq,Ivanov:2007de,Ginzburg:2008efb,Ginzburg:2009dp} 
and their partial neglecting is infamous for giving rise to 
errors in renormalization starting at two loops, see discussion in~\cite{Bednyakov:2018cmx} and references therein~(see also~\cite{Schienbein:2018fsw}). Non-canonical kinetic terms generally enter the beta functions at all levels. In the canonical choice of basis, this only becomes visible through the field anomalous dimensions, which is typically considered to be a subleading effect. However, the non-trivial transformation of the kinetic terms under goofy transformations (and rescalings) is an all-order effect that affects the beta functions at the non-perturbative level. Exposing the complete couplings of the kinetic terms, hence, would allow to track their all-order covariant behavior, which is obscured if one immediately chooses the canonical basis.

In the most general basis, the gauge-kinetic terms of the 2HDM scalars are exposed as
\begin{equation}\label{eq:mostGeneralKinetic}
\mathcal{L}_K=K_{ij}\left(D_\mu\Phi_i\right)^\dagger\!\left(D^\mu\Phi_j\right)\qquad\text{with}\qquad K_{ij}=K_{ji}^*\;.
\end{equation}
The dimensionless ``coupling coefficients'' $K_{ij}$ of the kinetic terms form a hermitean matrix $K$ and are closely related to the wave-function renormalization coefficients.

We briefly recap the steps that are done in order to arrive at canonical gauge-kinetic terms, which is usually implicit if one directly starts from the canonical basis. Like any hermitean matrix, $K$ can be written as 
\begin{equation}
 K~=~U^\dagger\mat{k_1 & 0 \\ 0 & k_2}U\,, 
\end{equation}
where $U$ here is a unitary $2\times2$ matrix, and $k_{1,2}\in\mathbbm{R}$ the real eigenvalues of $K$. Regarding the most general kinetic term above, the matrices $U$ can be absorbed in a \textit{regular} field redefinition of $\Phi$ and $\Phi^*$.
Obtaining unit diagonal entries is done by a subsequent rescaling (no sum)
\begin{align}
 \Phi_i=\sqrt{k_i}\,\Phi'_i\,, \qquad\text{and}\qquad \Phi_i^*=\sqrt{k_i}\,\Phi'^*_i\;.
\end{align}
Note that this rescaling is actually a goofy transformation if any one of the $k_i<0$. After these field redefinitions one arrives at the usual starting point of canonical kinetic terms in Eq.~\eqref{eq:kinetic_1}, together with the most general scalar potential in the canonical basis, Eq.~\eqref{eq:pot}.

Regular symmetry transformations, by definition, leave the kinetic term invariant in the canonical basis. This is a defining feature of regular transformations that holds in any basis. In a generic non-canonical basis this means that the transformation matrix of any regular transformation commutes with $K$. By contrast, goofy flavor- or CP-type transformations transform the hermitean matrix $K$ as $K\mapsto A^\mathrm{T} K\,B^*$, or $K\mapsto D^\dagger K^\mathrm{T}C$, respectively. Requiring invariance under such a goofy transformation gives constraints on the entries of the hermitean matrix $K$ that can be obtained from the first three columns of mass entries in Tab.~\ref{tab:Goofy_relations} by the formal replacement $m_{ij}\rightarrow K_{ij}$. Even if not enforced, the goofy transformation induces a non-trivial, all-order transformation behavior for the couplings of a theory, just like for any regular transformation but also including non-trivial transformation behavior of the entries of $K$. Hence, next to the beta functions of couplings and masses, our argument of the previous section can be applied in exactly the same manner for the full set of couplings of the gauge-kinetic terms, i.e.\ the hermitean matrix of wave function renormalization coefficients $K_{ij}$. Systematically deriving the effect of goofy transformations on the full system of RG equations would require to have them formulated for the most general $K$ to begin with. This would allow us to take the discussion beyond sign flips and investigate the full multiplet (triplet in this case) structural constraints on the beta functions (and anomalous dimensions) arising from goofy transformations. However, all currently available packages that compute higher-loop beta functions always make the (possible and not incorrect) choice to start from the canonical basis. Hence, we once more encourage a formulation in the most general case. As a side remark, we note that including the non-trivial kinetic terms should also give further insights for the recently greatly progressed formulation of RG running entirely in basis invariants~\cite{Bednyakov:2025sri}.

One should not confuse the entries of $K$ or its eigenvalues $k_i$ (or their signs) with the possible signs $\kappa_i$ of the goofy transformations in the canonical basis (see Eq.~\eqref{eq:LKtrafo}). The eigenvalues $k_i$ generally run under the RG and the points $k_i=0$ are a priori not any more special than setting any other of the couplings, e.g.\ $\lambda_i$'s, in the potential to zero. That is, zero eigenvalues of $K$ are generically lifted under RG running, which is a very well-known situation also regarding the hermitean mass terms (much of the following discussion also applies to mass terms but we focus on the kinetic terms here).

Nonetheless, there is an intuition that the points with zero eigenvalues are somehow special. This is substantiated by the existence of goofy transformations which can enforce such zero eigenvalues. If a sign-flipping goofy transformation acting on $k_i$ is respected by all but the gauge-kinetic terms, the points with $k_i=0$ cannot be crossed in the RG flow because they correspond to a situation with enhanced symmetry. But the nature of $k_i=0$ as (partial) fixed points requires special (goofy symmetric) values of other parameters. A zero crossing of one of the kinetic term eigenvalue \textit{does} seem to be possible for generic parameters. It would unavoidably happen if the absolute value of the off-diagonal element $|K_{12}|$ runs to a value that dominates the other elements.\footnote{%
The eigenvalues are given by $2\,k_{1,2}~=~K_{11}+K_{22}\pm\sqrt{(K_{11}-K_{22})^2+4|K_{12}|^2}\;.$
}
Completely in line with that, controlling the off-diagonal entries by a symmetry is only possible by a goofy transformation.

Obviously, for non-zero eigenvalues it is always possible to chose a basis in which the eigenvalues are positive (or even unity). Hence, if one perpetually restores this condition by rotating to a canonical basis for each incremental step in the RG running, one gets ``blind'' to the zero crossing of kinetic eigenvalues, as the running of propagator poles gets deferred to other couplings by the basis rotation and rescaling. Nonetheless, we are convinced that singular effects in the kinetic terms, even if scaled away to other sectors, should leave a physical signature on the whole theory, since physical statements cannot depend on choice of basis or scheme. 

Since a zero eigenvalue cannot be rescaled, it should correspond to a situation discontinuous from such perpetual rescaling. For the case of explicitly broken goofy symmetry, this is exactly the same discontinuity that is displayed by other RG fixed points for regular symmetries: One can get arbitrarily close to a fixed point in the RG flow, but the fixed points themselves can never be reached exactly. Hence, the goofy transformations are very important as they ensure that, at least from some directions in parameter space, the $k_i=0$ case indeed to corresponds to RG fixed points.

The fixed points with one or several $k_i=0$ and exactly preserved goofy symmetries corresponds to a special class of decoupling behavior. Note that fields with (asymptotically) vanishing gauge-kinetic terms do not entirely vanish from the theory, but correspond to a specific kind of classical background field (one may also call this an auxiliary field or mean field). In many situations discussed above, these fields also feature a vanishing mass gap.\footnote{%
Remarkably, goofy symmetric situations with CP-type symmetry that we have discussed, namely $\mathrm{CP1}_G$ and $\mathrm{CP2}_G$ still do involve non-vanishing mass parameters with two equal but opposite-sign eigenvalues. Correspondingly, these allow for non-vanishing, yet goofy symmetric kinetic terms with $k_1=-k_2\neq0$.} The interpretation as classical background fields is supported by the fact that rescaling small kinetic terms to unity corresponds to scaling the field value up, i.e.\ corresponds to large occupation numbers in a scheme where propagator residues have been rescaled to unity. After imposing the exact goofy symmetry, incl.\ setting the gauge-kinetic term to zero (or asymptotically reaching both conditions in the RG flow), the then-background fields will only enter in even powers for sign-flipping goofy transformations (or other invariant combinations if considering more complicated goofy transformations). That is, these fields will assume non-propagating background values that form singlets under goofy transformations and couple to the remaining fields in the theory in exactly such a way that the goofy transformation is respected. In this sense, the goofy-symmetric fixed points are hosts of new quantum theories in which the occurrence of classical background fields is naturally implemented from a parent quantum theory as a boundary condition. Since these points are never exactly reached from the parent theory, corrections around the strictly mean-field theory are computable in perturbation theory.

\section{Further comments}
\label{sec:comments}
It is interesting to note that goofy transformations have the tendency to prohibit bare mass terms, often even leading to entirely classically scale invariant theories. 
This is evidence of the fact that goofy transformations are closely related to rescaling transformations.
A proper phenomenological investigation of the theories discussed here, hence, should take into account the quantum corrected effective potential including also gauge and Yukawa couplings, as well as the possibility of dimensional transmutation \`a la Coleman-Weinberg. Even though zeros in the mass matrix originating from partially sign-flipping goofy transformations are generally lifted, we take the fact that bare mass terms are, in principle, controllable by goofy transformations as a strong hint that new interesting avenues for the electroweak hierarchy problem can be found using goofy transformations.

Likewise, the close connection between goofy and scale transformations should be further scrutinized. Not only should one start from the most general kinetic terms to investigate the all-order transformation of the beta functions under goofy transformations, but the same thing should be done for rescaling and more general conformal transformations, which would give additional all-order constrains on the beta functions~\cite{deBoer:2025xx}.
We think it is presently not excluded that this may allow for a full computation of the beta functions to all orders in a closed form.

Furthermore, we argue that goofy transformations could be a key missing ingredient for resolving the flavor puzzle: Goofy transformations can be extended to fermions and considered in the SM. For example, it seems interesting to investigate situations with goofy symmetric Yukawa couplings. Those would be radiatively broken (at least) by gauge corrections that are sensitive to relative signs of the gauge-kinetic terms. The sensitivity to generation-dependent relative signs may allow for new variations of scenarios with radiative mass generation. For example, this could offer a way to overcome the usual problem of either too large or too small mass vs.\ mixing hierarchies in radiative flavor models, see e.g.~\cite{Weinberg:2020zba}.

In supersymmetric theories of flavor, on the other hand,
non-canonical kinetic terms have also been discussed in applications to the SM flavor puzzle and strong CP problem~\cite{Hiller:2001qg,Hiller:2002um}. The non-canonical kinetic terms there originate from a non-canonical K\"ahler potential. Here, goofy flavor transformations should be of considerable relevance as they -- unlike ordinary regular discrete or modular flavor symmetries~(see e.g.~\cite{Feruglio:2017spp,Xing:2020ijf} and references therein) -- are able to directly constrain the K\"ahler potential. This is of great importance, as the major roadblock in making this class of theories predictive is the form of their K\"ahler potential which typically cannot be constrained by regular symmetries~\cite{Chen:2012ha,Chen:2013aya,Chen:2019ewa}. Goofy transformations can directly constrain the kinetic terms in full generality and, hence, can also be instrumental to constrain the non-canonical K\"ahler potential. 

We come back to the curious situation around the goofy symmetric RG fixed points. Regular outer automorphisms provide sufficient conditions to compute (partial) RG fixed hyperplanes to all orders~\cite{deBoer:2025xx} and these can be attractive, repulsive or separatrices. Goofy transformations, on the other hand, in addition seem to correspond to specific decoupling regimes, where one or several of the fields turn into non-propagating backgrounds fields. Approaching a goofy fixed point in the RG flow corresponds to a situation where fields with non-trivially transforming kinetic terms dynamically approach a quasi-classical regime. Sitting exactly at the goofy symmetric fixed point (which we recall, is never a limit reached by continuous RG evolution) corresponds to setting a field's gauge-kinetic term (and sometimes but not always mass) to zero ``by hand.''\footnote{%
In the non-canonical basis, it is clear that it is the wave function renormalization coefficient that breaks the symmetry, not itself the gauge coupling or the squared gauge couplings as these cannot do the required sign flip in a covariant manner, at least for the case of minimally gauge-invariantly coupled scalars.}
Since fields without propagators are usually considered to be classical background fields (or auxiliary fields, or mean fields) we think it is fair to call the situation of approaching the RG fixed point a situation of ``dynamical classicalization.''\footnote{%
For an earlier use of the term ``classicalization'' see~\cite{Dvali:2016ovn} and references therein.}
The QFT in such situations approaches a point where it automatically ``integrates out'' one or more of its fields in the RG flow and approaches the classical limit for this field. Exactly as expected from any quantum-to-classical transition, the classical limit here is not continuously connected to the quantum theory but corresponds to an asymptotic fixed point that can never be completely reached in the RG flow. In this sense, the points with vanishing kinetic terms and full goofy symmetry are isolated points in the parameter space of the theory.
If the kinetic terms are non-zero, no matter if positive or negative, they can always be brought to the canonical form by a field redefinition. The interpretation of approaching a semi-classical regime is missed in schemes where the propagator poles (i.e.\ wave function renormalization coefficients) are, by assumption, perpetually rescaled to some arbitrary value (for example, unity). While such a rescaling is always possible, it would obscure the behavior of the theory to approach this classical limit. Of course, no physical result can depend on the choice of basis (or scheme, in this context), which implies that even with strictly canonical terms, this regime must exist. Hence, the close-to singular behavior of kinetic terms in one scheme has to manifest itself by obscure behavior of other couplings in other schemes. For example, this suggest there could be a connection between goofy transformations in QED and the infamous Landau pole.

Finally, note that the most natural choice of running scale in cosmological settings is that of relative cosmological redshifts or, equivalently, temperature evolution of a free-streaming, i.e.\ propagating, background gas. The situation of ``dynamical classicalization'', here, corresponds to fields that get cosmologically redshifted into their goofy-symmetric fixed points. Not only would these assume a classical background value, but this background value would also dynamically decouple from further redshifts as the kinetic term vanishes. This explains how scalar vacuum expectation values, like the Higgs VEV, dynamically decouple from cosmological red-shifts and assume a non-redshifting mean field, while fields that keep non-zero propagators (including propagating classical ensembles like the CMB) continue to be subject to red-shift.
\section{Conclusions}
Transformations which are explicitly broken by the (gauge-)kinetic terms but would otherwise be good possible symmetries of a QFT are important, but have generally been overlooked in the past. This has changed with the seminal work of FGOO~\cite{Ferreira:2023dke}, who
discovered a previously unknown all-order RG fixed point of the 2HDM and related it to the first \textit{goofy} transformation. 

Here we have established a non-trivial transformation behavior of the kinetic term as a defining criterion for \textit{goofy} versus \textit{regular} transformations. We have provided a coherent treatment of regular and goofy transformations and their respective CP- and flavor-type subvariants. This allowed us to uncover several new goofy transformations of the 2HDM and, associated with them, new all-order RG stable fixed points.

In the simplest case, goofy transformations lead to sign-flips of one or several gauge-kinetic terms.
Naively, one may think this would unavoidably map physically well-defined theories to unphysical (perhaps unbounded from below) regimes, but this is not necessarily the case. As we have explicitly shown, it is never the sign of the kinetic term alone that determines the physicality of a theory, but always a combination of relative signs also involving the interactions. In particular, theories with wrong-signed gauge-kinetic terms can correspond to perfectly healthy theories, just written in an unusual basis. 

The main question we have addressed in this paper is the one after all-order stability of the parameter relations implied by goofy transformations. We have given a general argument for the all-order stability of regular and certain classes of goofy transformations. While we only sketched parts of the argument relevant here, a general formulation will be presented in the forthcoming~\cite{deBoer:2025xx}. Our argument is based on the outer automorphism nature of regular and goofy transformations. It allows to constrain in a very general manner the non-perturbative structure of beta functions and their associated fixed points, for all QFTs. Applied to the present case, our argument explains why global-sign-flipping goofy parameter relations are radiatively stable to all orders, even though they are explicitly broken by the gauge-kinetic terms. Furthermore, it allowed us to understand why goofy parameter relations based on transformations with only relative-sign-flipping behavior of the gauge-kinetic terms generally do receive quantum corrections, and of what type and at what order. We have emphasized that more general insights are possible if the RG running were not formulated starting from the canonical basis, but from the most general basis for the kinetic terms.

The fact that goofy transformations often lead to prohibition of mass parameters suggests a close connection between goofy and conformal-type symmetry transformations,
but their detailed relation remains to be investigated. In any case, goofy transformations can prohibit bare scalar mass terms, meaning that they are potentially of great interesting to tackle the hierarchy problem. Furthermore, goofy transformations can be interesting flavor symmetries, in which case they can control non-canonical kinetic terms, for example, stemming from the K\"ahler potential of supersymmetric theories. Radiative breaking of goofy relations by gauge interactions, furthermore, becomes sensitive to the \textit{relative} sign of the gauge coupling of different generations, which points to unexplored avenues in radiative mass generation.

Last but not least, approaching the RG fixed points of exact goofy symmetry corresponds to a situation where originally dynamical quantum fields are driven to become non-propagating classical background fields (auxiliary fields, or mean fields), a phenomenon that we have dubbed ``dynamical classicalization'' in QFT. Also this curious aspect of goofy transformations remains to be explored further.

While we have used the 2HDM as useful playground, our discussions applies to QFTs in general. Whether any larger goofy symmetries are actually at play (if the transformations are extended to imaginary-valued gauge fields or space-time as in~\cite{Ferreira:2023dke}), or if field-space complexifications as in~\cite{Haber:2025cbb} ultimately play a role to better understand goofy transformations, is certainly an interesting question that still deserves attention. However, here we have shown that the goofy relations and their RG stability can very well be understood from within the starting QFT itself, without requiring extensions. 

\enlargethispage{1cm}
Goofy transformations reveal an important and extremely deep relation to many fundamental questions in particle physics and QFTs in general. They are here to stay, and it will be important and interesting to gather more knowledge about them in all respects.
\section*{Acknowledgments}
I am grateful to Thede de Boer, Pedro M.\ Ferreira, Bohdan Grzadkowski, Howard E.\ Haber, Igor P.\ Ivanov, David E.\ Kaplan, Jeffrey Kuntz, Per Osland, Apostolos Pilaftsis, Jo\~ao P.\ Silva and Anders E.\ Thomsen for related exchanges and conversations. This work is supported in part by the Portuguese Funda\c{c}\~ao para a Ci\^encia e a Tecnologia (FCT) through project \href{https://doi.org/10.54499/2023.06787.CEECIND/CP2830/CT0005}{2023.06787.CEECIND} and contract \href{https://doi.org/10.54499/2024.01362.CERN}{2024.01362.CERN}, partially funded through POCTI (FE-DER), COMPETE, QREN, PRR, and the EU.

\appendix
\section{Appendix}\label{sec:appendix}
Possible choices for the explicit matrix generators of the regular transformations, defined in Eq.~\eqref{eq:RegularTrafos}, or goofy transformations as in Eq.~\eqref{eq:GoofyTrafos2HDM}, are given by
\begin{align}\notag
 &\mathbbm{Z}_{2,(G)}:& &S=\mat{1&0\\0&-1}\equiv\sigma_3\;,& &\text{CP}1_{(G)}:& &X=\mathbbm{1}_2\;,& \\
 &\text{U(1)}_{(G)}:& &S=\mat{e^{-i\xi}&0\\0&e^{i\xi} }\;,& &\text{CP}2_{(G)}:& &X=\mat{0&-1\\1&0}~\equiv~\varepsilon\;,& \\ \notag
 &\text{SU(2)}_{(G)}:& &S=\mat{e^{-i\xi}\cos\theta&-e^{-i\psi}\sin\theta\\ e^{i\psi}\sin\theta&e^{i\xi}\cos\theta }\;,& &\text{CP}3_{(G)}:& &X=\mat{\cos\theta&-\sin\theta\\\sin\theta&\phantom{-}\cos\theta }\;.& 
\end{align}
Here $\xi$, $\psi$, and $\theta$ are group parameters that are allowed to take arbitrary real values (up to some obviously excluded special values, which would make the transformations collapse to smaller transformations).
\bibliography{Bibliography}
\addcontentsline{toc}{section}{Bibliography}
\bibliographystyle{utphysM}
\end{document}